\documentclass[twocolumn,english,superscriptaddress]{revtex4}
\usepackage[T1]{fontenc}
\usepackage[latin9]{inputenc}
\usepackage{bm}
\usepackage{amstext}
\usepackage{amssymb}
\usepackage{graphicx}
\usepackage{esint}

\makeatletter

\providecommand{\tabularnewline}{\\}

\@ifundefined{textcolor}{}
{%
 \definecolor{BLACK}{gray}{0}
 \definecolor{WHITE}{gray}{1}
 \definecolor{RED}{rgb}{1,0,0}
 \definecolor{GREEN}{rgb}{0,1,0}
 \definecolor{BLUE}{rgb}{0,0,1}
 \definecolor{CYAN}{cmyk}{1,0,0,0}
 \definecolor{MAGENTA}{cmyk}{0,1,0,0}
 \definecolor{YELLOW}{cmyk}{0,0,1,0}
}

\makeatother

\usepackage{babel}
\begin{document}

\title{Critical Phenomena in Hyperbolic Space}

\author{Karim Mnasri}

\affiliation{Institute for Theory of Condensed Matter, Karlsruhe Institute of
Technology, 76131 Karlsruhe, Germany}

\affiliation{Institute of Theoretical Solid State Physics, Karlsruhe Institute
of Technology, 76131 Karlsruhe, Germany}

\author{Bhilahari Jeevanesan}

\affiliation{Institute for Theory of Condensed Matter, Karlsruhe Institute of
Technology, 76131 Karlsruhe, Germany}

\author{J\"org Schmalian}

\affiliation{Institute for Theory of Condensed Matter, Karlsruhe Institute of
Technology, 76131 Karlsruhe, Germany}

\affiliation{Institute for Solid State Physics, Karlsruhe Institute of Technology,
76131 Karlsruhe, Germany}
\begin{abstract}
In this paper we study the critical behavior of an $N$-component
$\bm{\phi}^{4}$-model in hyperbolic space, which serves as a model
of uniform frustration. We find that this model exhibits a second-order
phase transition with an unusual magnetization texture that results
from the lack of global parallelism in hyperbolic space. Angular defects
occur on length scales comparable to the radius of curvature. This
phase transition is governed by a new strong curvature fixed point
that obeys scaling below the upper critical dimension $d_{uc}=4$.
The exponents of this fixed point are given by the leading order terms
of the $1/N$ expansion. In distinction to flat space no order $1/N$
corrections occur. We conclude that the description of many-particle
systems in hyperbolic space is a promising avenue to investigate uniform
frustration and non-trivial critical behavior within one theoretical
approach. 
\end{abstract}
\maketitle

\section{Introduction}

\label{sec:intro} 
\begin{figure}[h]
\centering{}\includegraphics[width=0.95\columnwidth]{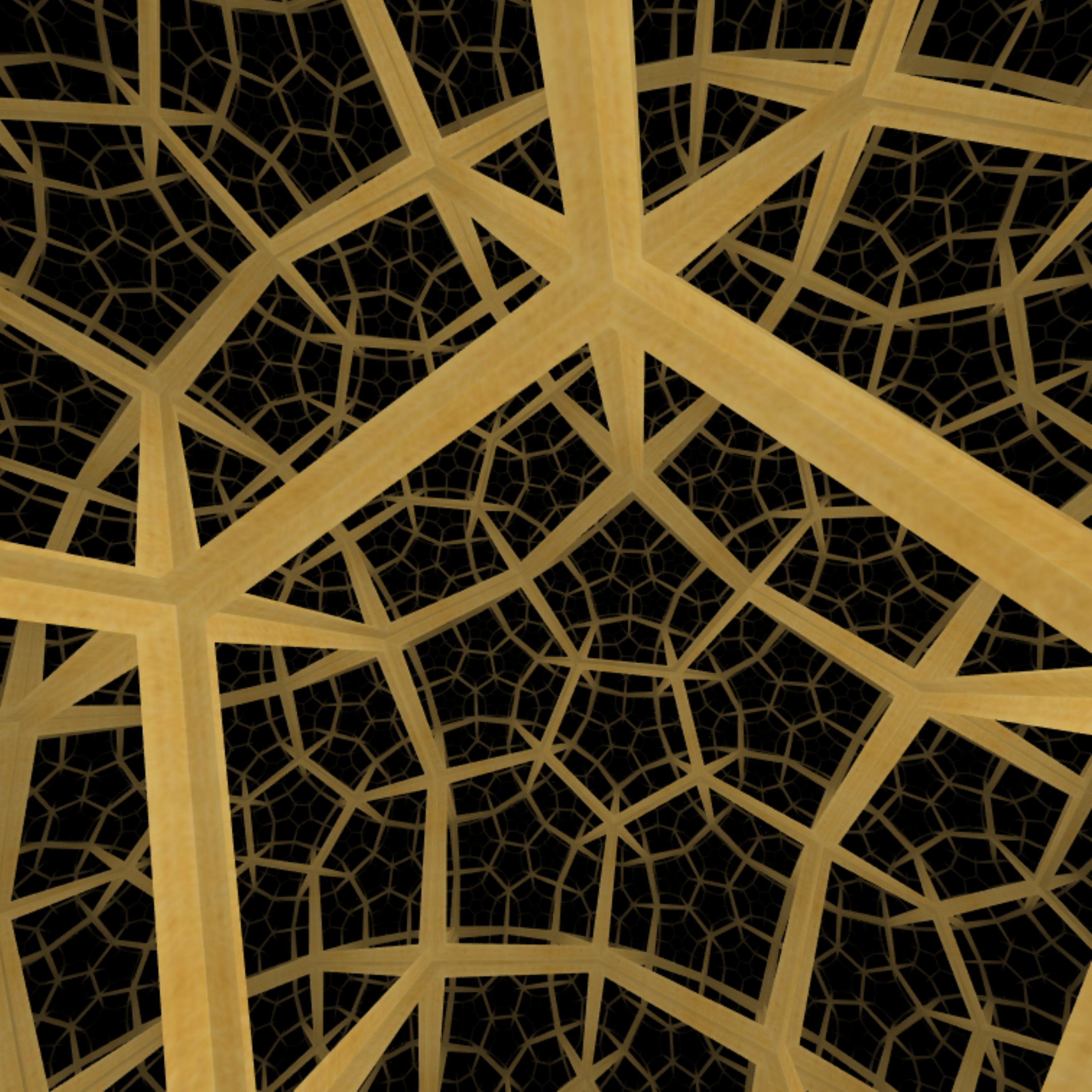}\protect\caption{\label{fig:hypTess} Tessellation of three-dimensional hyperbolic
space by dodecahedra (created with \cite{hyp})}
\end{figure}

Field theory and statistical mechanics in geometries
with negative curvature are of increasing interest. While a direct
application to the spacetime of our universe seems to require a positive
cosmological constant, a wide range of many-particle problems are
closely tied to problems with negative spatial curvature. For example,
field theories in hyperbolic space are increasingly studied because
of its direct relation to anti-de Sitter space. The latter is essential
for the duality between strong coupling limits of certain quantum
field theories and higher-dimensional gravity theories\cite{Maldacena98,Gubser98,Witten98}.
The scaling behavior near critical points in hyperbolic space, being
the Wick-rotated version of anti-de Sitter space, may therefore be
of relevance in the analysis of strong coupling theories. On the other
hand, networks like the Bethe lattice, that have been studied early
on in the statistical mechanics of phase transitions \cite{Kurata53,Domb60,Thorpe82}
and that have received renewed interest in the context of the dynamical
mean-field theory of correlated fermions\cite{RevModPhys.68.13,PhysRevB.71.235119},
quantum spin glasses\cite{PhysRevB.78.134424}, or bosons\cite{PhysRevB.80.014524},
can be considered as a regular tiling of the hyperbolic plane \cite{mosseri1982bethe}.
To be precise, if one considers a regular tiling $\left\{ p,q\right\} $,
where $p$ refers to the degree of a polygon and $q$ to the number
of such polygons around each vertex, then the Bethe lattice with coordination
number $q$ corresponds to $\left\{ \infty,q\right\} $. All regular
tilings of the hyperbolic plane with $\left(p-2\right)\left(q-2\right)>4$
are possible \cite{mosseri1982bethe}. Obviously the square lattice
$\left\{ 4,4\right\} $, the triangular lattice $\left\{ 3,6\right\} $,
and the honeycomb lattice $\left\{ 6,3\right\} $, i.e. the only possible
tilings with regular polygons of the two-dimensional flat space, are
just excluded. It was already stressed in Ref.\cite{mosseri1982bethe}
that hyperbolic tiling might be used to interpolate between the mean-field
behavior of the Bethe lattice and a lattice that might be close to
the square or honeycomb lattice. This may offer an alternative approach
to study corrections beyond dynamical mean-field theory. A tiling
of three-dimensional hyperbolic space with dodecahedra is shown in
Figure \ref{fig:hypTess} (see \cite{hyp}). \\
Finally, effects of uniform frustration are often captured in terms
of certain background gauge fields or by embedding a theory in curved
space \cite{0953-8984-17-50-R01,PhysRevLett.53.1947}. An interesting
case of tunable uniform frustration is found by studying a given flat-space
problem in curved space with inverse radius of curvature $\kappa$,
an idea that was introduced in \cite{nelson2002defects,PhysRevLett.50.982,PhysRevB.28.6377,PhysRevB.28.5515}.
 Here, the problem of packing identical discs was studied in a hyperbolic
plane. While in flat space packing in hexagonal close-packed order
is possible, in hyperbolic space this order is frustrated by the fact
that gaps open up between neighboring discs. This facilitated the
study of packing properties as a function of frustration. The latter
can be varied by changing the spatial curvature $\kappa$. Thus one
might capture packings that are not allowed in flat space, as it occurs
for clusters with icosahedral local order, in terms of a non-frustrated
model that is embedded in a curved geometry. These ideas were also
employed in studies of glass transitions in hyperbolic space (\cite{PhysRevLett.104.065701,PhysRevLett.101.155701}),
where the authors performed molecular dynamics simulations on the
hyperbolic plane for a Lennard-Jones liquid and found that the fragility
of the resulting glass is tunable by varying $\kappa$.\\
Given these applications of negatively-curved geometries, it is an
interesting question to ask how phase transitions of classical and
quantum models will behave in such curved spaces. Significant numerical
work has been devoted to studies of classical spin models in hyperbolic
space. The thermodynamic properties of Ising spins placed on the vertices
of lattices in hyperbolic space were studied in \cite{1751-8121-41-12-125001,PhysRevE.78.061119,PhysRevE.84.032103,PhysRevE.86.021105}.
In order to perform Monte Carlo simulations on finite two-dimensional
lattices a negatively curved background is created by tessellating
the hyperbolic plane with regular $n$-gons. All these works have
found the phase transition to follow mean-field behaviour. In particular,
various critical exponents were measured and found to numerically
coincide closely with mean-field exponents. One should, however, keep
in mind that the detailed protocol for measuring the critical exponents
in these works is somewhat different from the usual flat space protocol.
The mean-field behavior is supported by a Ginzburg criterion for $\phi^{4}$-theories
in hyperbolic space that was discussed in Ref. \cite{1742-5468-2015-1-P01002}.
There this behavior was rationalized by arguing that the Hausdorff
dimension of hyperbolic space is infinite. \\
The problem now arises to address the question of phase transitions
in three-dimensional hyperbolic space. In particular, it is natural
to ask, whether there is scaling as in flat $d$-dimensional space,
or if the transition is genuinely of mean-field type. If phase transitions
in hyperbolic space were of mean-field nature, it would imply for
systems that are below their upper critical dimension in flat space,
that an arbitrarily small curvature $\kappa$ would lead to a violation
of scaling. An alternative possibility is that scaling continues to
be valid in hyperbolic space with a new fixed point characterized
by new exponents. The above numerics would in that case indicate that
some exponents take their mean-field values. Because of hyperscaling
this cannot be the case for all critical exponents. In the case of
a new fixed point there are obvious questions: what is the universality
class and what are the critical exponents?\\
 \\
We answer these questions in this paper by studying analytically the
problem of an $N$-component continuum $\bm{\phi}^{4}$-theory in
three-dimensional hyperbolic space in a large-$N$ expansion. In the
discussion section we comment on the generalization to different dimensions.
Section \ref{sec:model} of this paper contains the exposition of
the $\bm{\phi}^{4}$ model in hyperbolic space. We find that the theory
possesses a second order phase transition, that scaling is obeyed
below the upper critical dimension of the flat space and that the
exponents are given by the leading order terms of the $1/N$ expansion.
To be specific, we find that the leading order $1/N$-corrections
to the exponents vanish. In addition we give general arguments that
support the conjecture that all higher order $1/N$-corrections should
vanish as well. The technical steps of our calculation are as follows:
Using the momentum space analysis of section \ref{MomentumSpace},
we locate the critical temperature of this phase transition and find
the magnetization texture of the ordered phase. In Section \ref{sec:transition}
we discuss the character of the phase transition and present the results
of the calculations of the exponents $\eta,\ \nu$ and $\gamma$ to
lowest order in $1/N$. In order to meaningfully identify the critical
point of this model, it is convenient to formulate the problems in
momentum space. As this representation in hyperbolic space does not
seem to exist in the condensed matter literature, we develop the necessary
parts in section IV. With this formalism it is now possible to deal
with the order $1/N$ correction to the critical exponents $\eta$
and $\gamma$. \\
We find that the exponents $\eta,\!\nu$ and $\gamma$ at lowest order
are those of three-dimensional flat space and not those of a mean-field
transition. However, in distinction to flat space, the $1/N$ corrections
vanish. The absence of higher-order corrections is found to be the
consequence of the finite curvature of hyperbolic space, which exponentially
cuts off fluctuations of wavelengths longer than the curvature radius.
This is in agreement with the general remarks on the regulating behavior
of hyperbolic space by Callan and Wilczek in \cite{Callan1990366}.\\
The critical exponents satisfy scaling and we discuss in the final
section how our results can be understood from the scaling of the
free energy in the presence of finite spatial curvature. \\
As a further result we calculated the magnetization texture of the
ordered state of this model. We find that uniform magnetization develops
in regions of size $1/\kappa$. Due to the lack of the concept of
global parallelism in hyperbolic space (\cite{PhysRevB.28.5515,PhysRevE.79.060106}),
these regions will necessarily be uncorrelated in their magnetization
direction.

\section{Model and Background Geometry}

\label{sec:model}

The model we are considering is an $N$-component $\bm{\phi}^{4}$-theory
given by the action 
\begin{eqnarray}
S & = & \int d^{3}x\sqrt{g}\ \frac{1}{2}\left[\mu_{0}\phi_{i}\cdot\phi_{i}+g^{\mu\nu}(\nabla_{\mu}\phi_{i})(\nabla_{\nu}\phi_{i})\right]\nonumber \\
 &  & +\int d^{3}x\sqrt{g}\ \frac{u}{4N}(\phi_{i}\cdot\phi_{i})^{2},\label{ModelAction}
\end{eqnarray}
with summation over $i,\mu$ and $\nu$ implied. Thus we are considering
here the three-dimensional version of $\bm{\phi}^{4}$-theory. Generalizations
to different dimensions are straightforward, as we discuss in the
final section. In the action, $g_{\mu\nu}$ is the metric of three-dimensional
hyperbolic space, which is a maximally symmetric space with negative
curvature, characterized by a single parameter, the curvature $\kappa$.
The quantity $g$ is the metric determinant and assures the proper
transformation property of the action. Hyperbolic space can be defined
as one of the two (equivalent) simply-connected three-dimensional
manifolds of points satisfying 
\begin{eqnarray}
x_{1}^{2}+x_{2}^{2}+x_{3}^{2}-x_{4}^{2}=-\frac{1}{\kappa^{2}}
\end{eqnarray}
inside Minkowski space. The coordinates $x_{i}$ are the cartesian
coordinates of Minkowski space. To derive a more convenient formulation,
the points may be parametrized by 
\begin{eqnarray}
x_{1} & = & \frac{1}{\kappa}\sinh\kappa r\sin\theta\cos\phi\\
x_{2} & = & \frac{1}{\kappa}\sinh\kappa r\sin\theta\sin\phi\\
x_{3} & = & \frac{1}{\kappa}\sinh\kappa r\cos\theta\\
x_{4} & = & \frac{1}{\kappa}\cosh\kappa r.
\end{eqnarray}
Minkowski space has a metric that is given by 
\begin{eqnarray}
ds^{2}=dx_{1}^{2}+dx_{2}^{2}+dx_{3}^{2}-dx_{4}^{2}.
\end{eqnarray}
This induces an intrinsic metric on the hyperbolic space with line-element
\begin{eqnarray}
ds^{2}=dr^{2}+\frac{1}{\kappa^{2}}\sinh^{2}\kappa r\left(d\theta^{2}+\sin^{2}\theta d\phi^{2}\right).
\end{eqnarray}
In the limit $\kappa\rightarrow0$ we regain three-dimensional flat
space. An additional length scale, $1/\kappa$, is present in hyperbolic
space, which is ultimately responsible for the non-trivial magnetization
texture that we derive below. Note that our results can straightforwardly
be applied to quantum phase transitions in hyperbolic space, if one
of the spatial coordinates is considered as imaginary time after the
usual Wick rotation.

\section{Phase transition and magnetization texture}

\label{sec:transition}

\begin{figure}[b]
\includegraphics[width=1\linewidth]{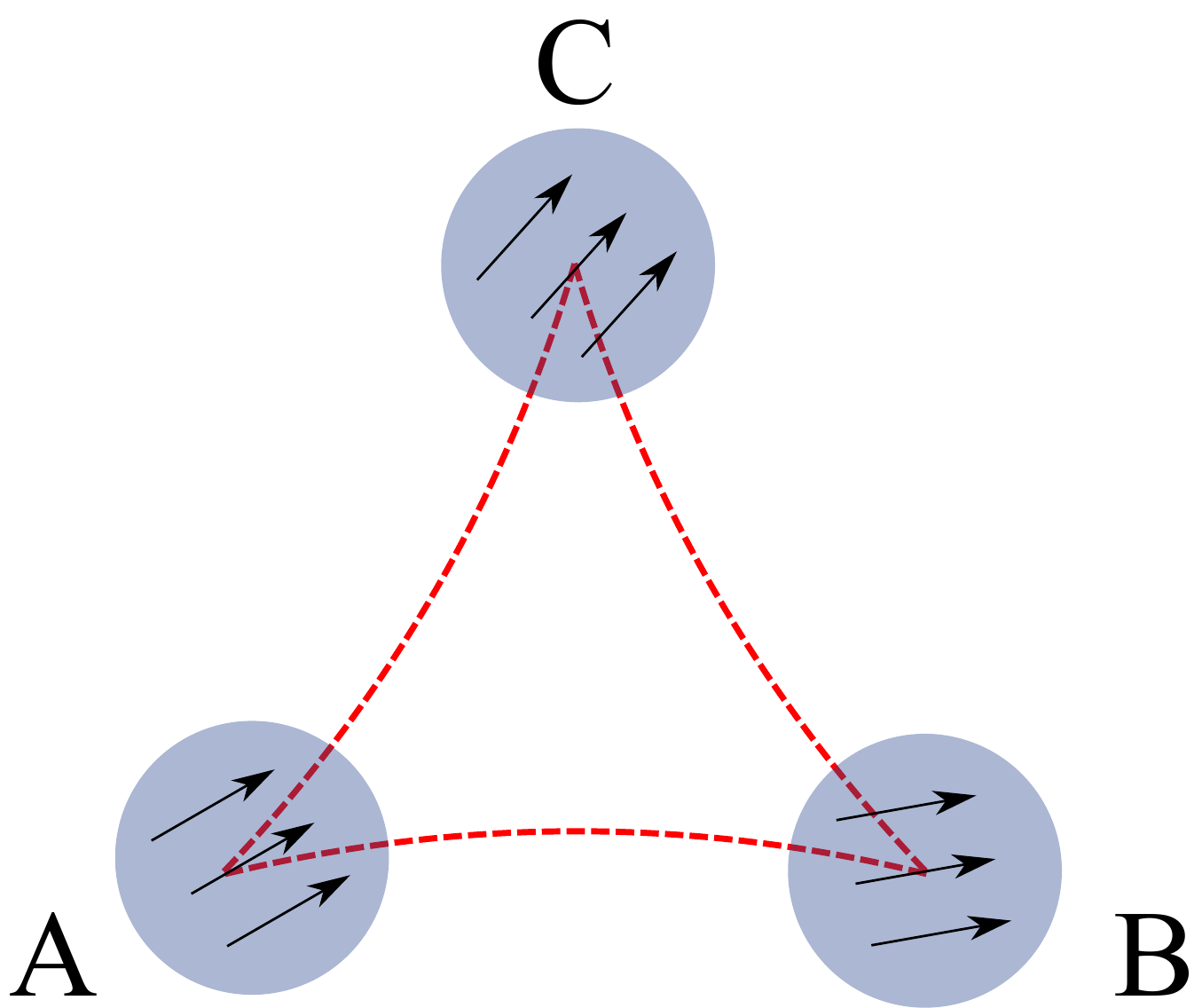} \protect\caption{\label{HypTriangle} Three magnetized patches in a plane, forming
a hyperbolic triangle. It is not possible to give a meaningful definition
of the direction of magnetization, since parallel-transport of a vector
from A to B and A to C, will not result in vectors that will match
upon parallel-transport from B to C or vice versa.}
\end{figure}

In three-dimensional flat space the model that we consider is known
to possess a second order phase transition, where the ordered state
corresponds to a symmetry-broken phase with uniaxial magnetization.
Geometrically this is not possible in hyperbolic space (\cite{PhysRevB.28.5515,PhysRevE.79.060106}),
since here a global direction is not a well-defined concept. Consider,
as shown in Fig. \ref{HypTriangle}, three locally magnetized patches
$A,B,C$, which are the corners of a hyperbolic planar triangle. 
 The direction of the order-parameter at $A$ may be parallel-transported
to $B$ and $C$ along the geodesics $\overline{AB}$ and $\overline{AC}$,
respectively. If now we continue the parallel-transport from $B$
to $C$ along $\overline{BC}$, the two transported magnetization
directions will not match. Instead, there will be an angular defect
$\theta$ between the two magnetizations that is proportional to the
enclosed area $\mathcal{A}$ of the hyperbolic triangle: 
\begin{eqnarray}
\theta=\mathcal{A}\kappa^{2}\label{eq:angDefect}
\end{eqnarray}
This formula follows from the fact that the vectors are parallel-transported
such that the angle between the geodesic curve and the vector is a
constant. Since hyperbolic triangles have angles which sum to $\pi-\mathcal{A}\kappa^{2}$
(\cite{coxeter1969introduction}), we are left with the angular defect
stated in (\ref{eq:angDefect}). Thus the ordered state in hyperbolic
space will in general be more complicated. Inside the radius of curvature,
where $\mathcal{A}<1/\kappa^{2}$, a uniform direction may be meaningfully
defined. \\
In order to determine the nature of the ordered state, we study the
symmetry-broken state of the action $S$ at the lowest order in a
$1/N$ expansion, i.e. we peform a saddle point analysis of the partition
function 
\begin{eqnarray}
Z=\int D\bm{\phi}\exp(-S),
\end{eqnarray}
and then include higher-order fluctuations in a systematic fashion.
Here, the action $S$ is given by: 
\begin{eqnarray}
S=\beta\int dV\ \frac{1}{2}\left[\bm{\phi}\left(\mu_{0}-\nabla^{2}\right)\bm{\phi}+\frac{u}{2N}\left(\bm{\phi}\cdot\bm{\phi}\right)^{2}\right].\label{eq:action}
\end{eqnarray}
We rewrite this by performing a Hubbard-Stratonovich decoupling of
the $(\bm{\phi}\cdot\bm{\phi})^{2}$ term, whereupon the action becomes
\begin{eqnarray*}
S=\frac{\beta}{2}\int dV\left[\bm{\phi}\left(\mu_{0}+i\lambda(\bm{x})-\nabla^{2}\right)\bm{\phi}+\frac{N}{2u}\lambda^{2}(\bm{x})\right].
\end{eqnarray*}
Now we integrate out all $\phi_{i}$ with the exception of the one
component, along which the spins near a chosen point order and which
we will label $\sigma(\bm{x})$. Moreover, we introduce a source field
$h(\bm{x})$ for $\sigma(\bm{x})$. This leads to the action 
\begin{eqnarray*}
S & = & \beta\int dV\ \frac{1}{2}\left[\sigma(\bm{x})\left(\mu_{0}+i\lambda(\bm{x})-\nabla^{2}\right)\sigma(\bm{x})+\frac{N}{2u}\lambda^{2}(\bm{x})\right]\\
 &  & +\frac{N-1}{2}\mathrm{Tr}\left[\log(\mu_{0}+i\lambda(\bm{x})-\nabla^{2})\right]\\
 & &-\beta\int dV\ h(\bm{x})\sigma(\bm{x})\\
\end{eqnarray*}
for the partition function $Z=\int D\sigma D\lambda\exp(-S)$.\\
 The saddle point solutions are determined by the conditions $\frac{\delta S}{\delta\sigma(\bm{x})}=0$
and $\frac{\delta S}{\delta\lambda(\bm{x})}=0$, which result in the
two equations 
\begin{eqnarray}
\left(\mu(\bm{x})-\nabla^{2}\right)\sigma(\bm{x}) & = & h(\bm{x})\label{eq:saddle_point1}\\
\mu(\bm{x})-\mu_{0} & = & uT\langle\bm{x}|\frac{1}{\mu(\bm{x})-\nabla^{2}}|\bm{x}\rangle\nonumber \\
 &  & +\frac{u}{N}\sigma^{2}(\bm{x})\label{eq:saddle_point2}
\end{eqnarray}
with $\mu(\bm{x})=\mu_{0}+i\lambda(\bm{x})$.\\
We will use these equations to work out the critical exponents including
higher-order corrections in section IV. Here, we will only analyze
equation (\ref{eq:saddle_point1}) to find the susceptibility $\chi(\bm{x},\bm{x}')=\delta\sigma(\bm{x})/\delta h(\bm{x}')|_{h(\bm{x}')=0}$,
which, by virtue of (\ref{eq:saddle_point1}), satisfies 
\begin{eqnarray}
\left(\mu(\bm{x})-\nabla^{2}\right)\chi(\bm{x},\bm{x}') & = & \delta(\bm{x}-\bm{x'}).
\end{eqnarray}
As we approach the phase transition from high temperatures, $\mu(\bm{x})$
may be assumed to be homogeneous. Using the formalism of the next
section, this equation may be transformed into momentum space whereupon
it becomes 
\begin{equation}
\chi_{p}=\frac{1}{\mu+\kappa^{2}+p^{2}},\label{chiMomentum}
\end{equation}
where $p\geq0$ are the eigenvalues of the Laplacian. Note the presence
of the `mass term' $\kappa^{2}$ in the denominator. This is a consequence
of the fact that the Laplace operator in hyperbolic space has a gapped
eigenvalue spectrum, as will be seen below. A criterion for the presence
of the phase transition is the condition that $\chi_{p}$ diverge.
The highest value of $\mu$ when this happens is $\mu=-\kappa^{2}$,
where the $p=0$ mode of the susceptibility diverges. Thus the phase
transition takes place at $\mu=-\kappa^{2}$ with an order that is
determined by the $p=0$ Fourier mode. In contrast to flat space,
the $p=0$ eigenmode of the Laplace operator cannot be one of homogeneous
order, in agreement with the foregoing argument about angular defects.
Instead, it corresponds to a diminishing of the magnetization $\sigma$
along the one direction, that we chose not to integrate out. In other
words, due to the lack of a global direction of magnetization, focussing
on one component of the $N$-component vector, entails that one is
eventually considering projections of the magnetization vector instead
of the full vector. The diminishing of this projection takes place
according to the formula 
\begin{eqnarray}
\sigma(r)=\sigma_{0}\frac{\kappa r}{\sinh\kappa r},\label{eq:p0mode}
\end{eqnarray}
where $r$ is the geodesic distance from the origin, where the unintegrated
component and local magnetization direction coincide and $\sigma_{0}$
is the magnitude of the magnetization at the origin. The phase transition
corresponds to the formation of infinitely many patches, more precisely
three-dimensional regions, of characteristic sizes $1/\kappa$, which
have nearly uniform magnetization. The decay of $\sigma(r)$ in Eq.
(\ref{eq:p0mode}) does not imply a decay of the magnitude of the
order parameter, but must be interpreted as the order parameter rotating
away from the chosen direction of the vector ${\bm{\phi}}$.

\section{Momentum space representation}

\label{MomentumSpace}

We come now to the technical part of this paper that will allow us
to analyze the saddle point equations (\ref{eq:saddle_point1}), (\ref{eq:saddle_point2})
and compute critical exponents. In order to make progress with the
calculations, it is convenient to obtain the momentum space representation
of functions that are translationally invariant in hyperbolic space.
Let $\psi(d_{PQ})$ be a given function of the geodesic distance between
two points $P$ and $Q$. The functional dependence on the two points
will not have an arbitrary form, but will rather be expressed through
the geodesic distance $d_{PQ}$ between these two points. This distance
is the length of the geodesic curve connecting these points. Explicit
computation of this length yields the formula 
\begin{eqnarray}
\cosh{\kappa d_{PQ}} & = & \cosh{\kappa r}\cosh{\kappa r'}-\sinh{\kappa r}\sinh{\kappa r'}\cos{\gamma}\nonumber \\
\cos{\gamma} & = & \cos{\theta}\cos{\theta'}+\sin{\theta}\sin{\theta'}\cos{(\phi-\phi')}.
\end{eqnarray}
The fact that such a function $\psi$ depends on the six coordinates
not in an arbitrary way, but only through the geodesic distance, allows
us to expand $\psi(d_{PQ})$ in terms of the eigenstates of the Laplace
operator in hyperbolic space. Since hyperbolic space may be defined
as the set of all points equidistant from the origin in Minkowski
space, this Laplace operator is identical to the one obtained by writing
down the 4-dimensional Laplace operator in angular coordinates and
restricting the distance from the origin to a constant. A similar
situation was considered by Fock \cite{Fock:1935kq}, who studied
the problem on a 3-sphere embedded in 4-dimensional euclidean space.
The eigenfunctions of the Laplacian on this 3-sphere are the generalized
spherical harmonics of three angles. Their full description was given
in \cite{Fock:1935kq}. We find the eigenfunctions of the Laplacian
in hyperbolic space by multiplying one of the angles in Fock's solution
by the imaginary unit, a prescription sketched briefly in an appendix
of \cite{lifshitz1963investigations}. \\
 The hyperbolic Laplacian is given by 
\begin{eqnarray}
\Delta & = & \frac{1}{\sinh^{2}\kappa r}\partial_{r}(\sinh^{2}\kappa r\partial_{r}\psi)+\frac{\kappa^{2}}{\sinh^{2}(\kappa r)}\Delta_{S^{2}}\nonumber \\
\Delta_{S^{2}} & = & \frac{1}{\sin\theta}\partial_{\theta}\left(\sin\partial_{\theta}\psi\right)+\frac{1}{\sin^{2}\theta}\partial_{\phi}^{2}\psi.
\end{eqnarray}
The eigenfunctions are then given by 
\begin{eqnarray}
\psi_{qlm}(r,\theta,\phi)=\Pi_{ql}(\kappa r)Y_{lm}(\theta,\phi)
\end{eqnarray}
with eigenvalues 
\begin{eqnarray}
\Delta\psi_{qlm}(r,\theta,\phi)=-(\kappa^{2}+q^{2})\psi_{qlm}(r,\theta,\phi).
\end{eqnarray}
Here the $Y_{lm}$ are the ordinary spherical harmonics on the 2-sphere
and the $\Pi_{pl}$ are special functions that solve the radial part
of the eigenvalue equation \begin{widetext} 
\begin{eqnarray}
\frac{d^{2}}{dr^{2}}\Pi_{ql}+2\kappa\coth{\kappa r}\frac{d}{dr}\Pi_{ql}-\frac{l(l+1)\kappa^{2}}{\sinh^{2}{\kappa r}}\Pi_{ql}(\kappa r)=-(\kappa^{2}+q^{2})\Pi_{ql}(\kappa r).
\end{eqnarray}
\end{widetext} The solutions can be expressed in a Rayleigh-type
formula 
\begin{eqnarray}
\Pi_{pl}(x)=\frac{\sinh^{l}x}{M_{l}}\left(\frac{d^{l+1}}{d(\cosh x){}^{l+1}}\right)\cos(px)
\end{eqnarray}
where 
\begin{eqnarray}
M_{l}^{2}=\left(\frac{q}{\kappa}\right)^{2}\left[\left(\frac{q}{\kappa}\right)^{2}+1^{2}\right]\dots\left[\left(\frac{q}{\kappa}\right)^{2}+l^{2}\right]
\end{eqnarray}
is a normalization constant. The differential equation being of the
Sturm-Liouville form, these functions satisfy the orthogonality relation
\begin{eqnarray}
\int\limits _{0}^{\infty}dr\sinh^{2}{(\kappa r)}\ \Pi_{ql}(\kappa r)\Pi_{q'l}(\kappa r)=\frac{\pi}{2}\delta(q-q').
\end{eqnarray}

\subsection{Addition theorem}

The eigenstates $\psi_{plm}(r,\theta,\phi)$ satisfy an addition theorem,
which was derived for the 3-sphere by Fock \cite{Fock:1935kq}. In
the latter case of the 3-sphere this formula is fully analogous to
the addition theorem for two-dimensional spherical harmonics. Again,
by multiplying one of the angles by the imaginary unit, we obtain
the corresponding addition theorem valid in hyperbolic space \begin{widetext}
\begin{eqnarray}
\frac{\sin(qd)}{\sinh(\kappa d)}=\frac{\kappa}{q}\sum_{l=0}^{\infty}(2l+1)\Pi_{ql}(\kappa r)\Pi_{ql}(\kappa r')P_{l}(\cos\gamma),\label{main_identity}
\end{eqnarray}
\end{widetext} where the $P_{l}(\cos\gamma)$ are the Legendre polynomials
in $\cos\gamma$. \\
As a demonstration of the use of this formula, let us derive the magnetization
texture of the $p=0$ eigenmode given in (\ref{eq:p0mode}). The eigenbasis
expansion of $\chi$ reads 
\begin{equation}
\chi(\bm{r},\bm{r}')=\int dp\sum_{l}(2l+1)\chi_{p}\Pi_{pl}(\kappa r)\Pi_{pl}(\kappa r')P_{l}(\cos\gamma).
\end{equation}
Insertion of $\chi_{p}$ from (\ref{chiMomentum}) into this equation
at the critical point $\mu=-\kappa^{2}$, yields the real-space form
of $\chi$. Now we construct the real-space form of only the $p=0$
mode, which gives 
\begin{eqnarray}
\sigma(d(\bm{r},\bm{r}')) & = & \lim_{p\rightarrow0}\sum_{l}(2l+1)\frac{\sigma_{p}}{p^{2}}\Pi_{pl}(\kappa r)\Pi_{pl}(\kappa r')P_{l}(\cos\gamma)\nonumber \\
 & = & \sigma_{0}\frac{\kappa d(\bm{r},\bm{r}')}{\sinh\kappa d(\bm{r},\bm{r}')},
\end{eqnarray}
as claimed.

\subsection{Extraction of coefficients and inversion formula}

The identity (\ref{main_identity}) will be crucial in obtaining the
expansion coefficients of a given function $\psi(d)$ of the geodesic
distance. This distance being a non-negative quantity, the value of
$\psi$ for negative arguments is irrelevant. In particular we may
redefine $\psi$ for negative arguments such that it becomes an even
function. This allows us to Fourier expand $\psi$ as follows 
\begin{eqnarray}
\psi(d)\sinh\kappa d & = & \frac{\kappa}{2\pi^{2}}\int\limits _{-\infty}^{+\infty}dp\ \psi_{p}p\sin({pd}),\label{eq:fourier_rep}
\end{eqnarray}
where we have chosen to split off a factor of $\kappa p/(2\pi^{2})$
in the definition of the expansion coefficient for later convenience.
Inserting (\ref{main_identity}) we obtain \begin{widetext} 
\begin{eqnarray}
\psi(d)=\frac{\kappa}{2\pi^{2}\sinh\kappa d}\int\limits _{-\infty}^{+\infty}dp\ \psi_{p}p\sin({pd})=\frac{\kappa^{2}}{2\pi^{2}}\int\limits _{-\infty}^{+\infty}dp\sum_{l=0}^{\infty}(2l+1)\psi_{p}\Pi_{pl}(\kappa r)\Pi_{pl}(\kappa r')P_{l}(\cos\gamma)\label{expansion}
\end{eqnarray}
\end{widetext} and have thereby managed to expand the arbitrary function
$\psi$ in the new basis with coefficients 
\begin{eqnarray}
\psi_{p}=\frac{\pi i}{\kappa p}\int\limits _{-\infty}^{+\infty}dx\psi(|x|)\sinh(\kappa x)e^{-ipx}.\label{eq:fourierComp}
\end{eqnarray}
Let us briefly comment on the structure of the expansion. Note that
in (\ref{expansion}) the expansion coefficient $\psi_{p}$ has no
dependence on $l$. In fact, the statement of (\ref{expansion}) is
that any function that depends on the set of coordinates $(r,\theta,\phi),(r',\theta',\phi')$
only through the geodesic distance of the two points, can have no
explicit $l$ or $m$ dependence of $\psi_{p}$. \\
 Conversely, to find the real-space function $\psi(d)$ from the knowledge
of the coefficients $\psi_{p}$ in the expansion (\ref{expansion})
we use (\ref{eq:fourier_rep}) which results in the inversion formula
\begin{eqnarray}
\psi(d)=\frac{\kappa}{2\pi^{2}i\sinh\kappa d}\int\limits _{-\infty}^{+\infty}dp\ \psi_{p}pe^{ipd}.\label{eq:inversion_formula}
\end{eqnarray}

\subsection{Convolution theorem}

Let $f$ and $g$ be two-point functions that depend on the geodesic
distances $d_{PQ}$ and $d_{QR}$, respectively. When we multiply
these functions and integrate $Q$ over all of hyperbolic space, the
resulting function $h$ can only depend on the geodesic distance between
points $P$ and $R$. This convolution will in general be difficult
to carry out in real-space. The fact that the $\Pi_{plm}$ and the
$Y_{lm}$ are orthogonal functions, however, allows us to reduce the
convolution of $f$ and $g$ to a multiplication in momentum-space.
This is seen explicitly by rewriting the relation 
\begin{eqnarray}
\int dV_{Q}f(d_{PQ})g(d_{QR})=h(d_{PR})
\end{eqnarray}
in the momentum representation (\ref{expansion}) with expansion coefficients
$f_{p},g_{p},h_{p}$ and using the orthogonality relations for the
radial functions and the spherical harmonics. Then this convolution
formula translates into 
\begin{eqnarray}
f_{p}g_{p}=h_{p}.\label{convolution_equation}
\end{eqnarray}
\\
 The solution of the Dyson equation below will require knowledge about
the momentum-space representation of the Dirac $\delta$-function
in hyperbolic space, which we denote by either $\delta_{PQ}$ or $\delta(\bm{r},\bm{r}')$.
We define this function by the condition that convolution of an arbitrary
function $\psi(d_{PQ})$ with $\delta(d_{QR})$ must yield $\psi(d_{PR})$.
Translating this condition into momentum space, we immediately read
off from (\ref{convolution_equation}) the relation $\delta_{p}=1$
and obtain thereby 
\begin{eqnarray}
\delta_{PQ} & = & \frac{\kappa^{2}}{2\pi^{2}}\int\limits _{-\infty}^{+\infty}dp\sum_{l=0}^{\infty}(2l+1)\nonumber \\
 &  & \times\Pi_{pl}(\kappa r)\Pi_{pl}(\kappa r')P_{l}(\cos\gamma).\label{deltaRep}
\end{eqnarray}
\\
 Conversely, however, the multiplication of two functions in real-space
does not translate into a simple convolution integral in momentum-space,
but rather a double-integral. Given the product 
\begin{eqnarray}
f(d_{PQ})g(d_{PQ})=h(d_{PQ})
\end{eqnarray}
the corresponding momentum-space equation is found by employing the
representation (\ref{eq:inversion_formula}) 
\begin{eqnarray}
h_{k}=\frac{1}{4\pi^{2}k}\int\limits _{-\infty}^{\infty}dp\int\limits _{-\infty}^{\infty}dqf_{p}g_{q}pq\tanh\left[\frac{p+q-k}{2\kappa}\pi\right]\label{eq:finMomConv}
\end{eqnarray}
i.e. instead of a single integral a double integral with kernel is
obtained. \\
 In the limit $k\rightarrow0$ the symmetry properties of $f_{p}$
and $g_{q}$ may be used to rewrite this kernel as 
\begin{eqnarray}
\lim_{k\rightarrow0}h_{k}=\frac{1}{8\pi\kappa}\int\limits _{-\infty}^{\infty}dp\int\limits _{-\infty}^{\infty}dq\ \frac{pqf_{p}g_{q}}{\cosh^{2}\frac{\pi(p-q)}{2\kappa}}.\label{eq:zeroMomConv}
\end{eqnarray}
This formula will be used in section \ref{sec:ExponentsCorrections}
in evaluating the corrections to the critical exponent $\gamma$.

\section{Critical exponents}

\label{sec:exponents} The previous formalism may now be employed
to analyze the saddle point equations (\ref{eq:saddle_point1}) and
(\ref{eq:saddle_point2}). These equations describe the physics of
the model at lowest order in $1/N$. Given this fact, all exponents
that we derive below represent the lowest order contribution in $1/N$.
In principle these are only the lowest order terms in an expansion
of the critical exponents in a power series in $1/N$. In flat three-dimensional
space there are indeed further corrections. The main result of this
paper, derived in section \ref{sec:ExponentsCorrections}, is to establish
the absence of such corrections for $\eta$ and $\gamma$ in three-dimensional
hyperbolic space. \\
 We begin by computing the bare Green's function $G^{(0)}(p)$ in
momentum space for constant $\mu$ in the action for $\sigma(\bm{x})$.
The real-space definition of the bare $G^{(0)}(\bm{r})$ is obtained
by inverting the action, i.e. 
\begin{eqnarray}
\left(\mu-\nabla^{2}\right)G^{(0)}(\bm{r},\bm{r}')=\delta(\bm{r},\bm{r}').
\end{eqnarray}
Inserting 
\begin{eqnarray}
G^{(0)}(\bm{r},\bm{r}') & = & \frac{\kappa^{2}}{2\pi^{2}}\int\limits _{-\infty}^{\infty}dp\sum_{l}(2l+1)\nonumber \\
 &  & \times G^{(0)}(p)\Pi_{pl}(\kappa r)\Pi_{pl}(\kappa r')P_{l}(\cos\gamma)
\end{eqnarray}
and the representation of $\delta(\bm{r},\bm{r}')$ in (\ref{deltaRep}),
it is found that 
\begin{eqnarray}
G^{(0)}(p)=\frac{1}{(\mu+\kappa^{2}+p^{2})}.\label{eq:bareGreen}
\end{eqnarray}
At the critical point, $\mu=-\kappa^{2}$, we have a power-law dependence
on $p$. Employing the inversion formula, we find the real-space dependence
\begin{eqnarray}
G^{(0)}(\bm{r},\bm{r}') & = & \frac{\kappa}{2\pi^{2}i\sinh\kappa d}\int\limits _{-\infty}^{\infty}dp\frac{pe^{ipd}}{(\mu+\kappa^{2}+p^{2})}\nonumber \\
 & = & \frac{\kappa}{2\pi}\frac{e^{-\sqrt{\mu+\kappa^{2}}d}}{\sinh\kappa d},\label{eq:bareGreenReal}
\end{eqnarray}
where $d$ is the geodesic distance between $\bm{r}$ and $\bm{r}'$.
Evidently, even at the critical point the Green's function decays
exponentially, in accord with the previous remarks about parallel-transport.
\begin{table}[b!]
\centering %
\begin{tabular}{c|c|c}
\hline 
 & $\eta$  & $\gamma$\tabularnewline
\hline 
$\kappa=0$  & $\frac{8}{3\pi^{2}}\frac{1}{N}+\mathrm{O}\left(\frac{1}{N^{2}}\right)$  & $2-\frac{24}{\pi^{2}}\frac{1}{N}+\mathrm{O}\left(\frac{1}{N^{2}}\right)$ \tabularnewline
$\kappa\neq0$  & $0$  & $2$ \tabularnewline
\hline 
\end{tabular}\protect\caption{Critical exponents in flat and hyperbolic space.}
\label{tablesymm} 
\end{table}

\subsection{Critical Exponent $\eta$}

Let us now proceed to the evaluation of the exponent $\eta$. At the
critical point $\mu=-\kappa^{2}$ the power-law form of the curved-space
bare Green's function in (\ref{eq:bareGreen}) agrees with the flat-space
limit. The exponent $\eta$ may therefore be defined through the relation
\begin{eqnarray}
G(p)\propto\frac{\Lambda^{-\eta}}{p^{2-\eta}}.\label{eq:defEta}
\end{eqnarray}
We see that the bare $G^{(0)}(p)$, i.e. the lowest order form of
the Green's function in an $1/N$ expansion, has $\eta=0$.

\subsection{Critical Exponents $\nu$ and $\gamma$}

\label{subsec:nuGamma} We now study the behavior of the correlation-length
as the critical temperature is approached from the disordered regime
by examining the saddle point equation (\ref{eq:saddle_point2}).
In the regime $\mu>-\kappa^{2}$, the magnetization will be zero.
We may therefore set $\sigma=0$ and obtain 
\begin{eqnarray}
\mu & = & \mu_{0}+\frac{uT}{2\pi^{2}}\int\limits _{0}^{\Lambda}dq\frac{q^{2}}{\mu+\kappa^{2}+q^{2}}\nonumber \\
 & = & \mu_{0}+\frac{uT}{2\pi^{2}}\Lambda-\frac{uT}{2\pi^{2}}\sqrt{\mu+\kappa^{2}}\pi.\label{eq:nu1}
\end{eqnarray}
At $T=T_{c}$, where $\mu=-\kappa^{2}$, we have 
\begin{eqnarray}
-\kappa^{2}=\mu_{0}+\frac{uT_{c}}{2\pi^{2}}\Lambda,
\end{eqnarray}
which allows us to remove $\mu_{0}$ from (\ref{eq:nu1}). Now close
to $\mu\gtrsim-\kappa^{2}$, the quantity $\sqrt{\mu+\kappa^{2}}$
dominates over $\mu+\kappa^{2}$. Thus we neglect the latter and find
\begin{eqnarray}
\sqrt{\mu+\kappa^{2}} & = & \frac{1}{\pi}\frac{T-T_{c}}{T_{c}}\Lambda.\label{eq:xiNearTc}
\end{eqnarray}
The length $\xi$ that diverges at the critical point is defined by
\begin{eqnarray}
\xi=\frac{1}{\sqrt{\mu+\kappa^{2}}},\label{eq:xidef}
\end{eqnarray}
which is the natural length scale of the problem as follows from the
correlation function Eq.(\ref{eq:bareGreen}) in its eigenbasis. At
the same time, the real-space form of the correlation function in
Eq.(\ref{eq:bareGreenReal}) shows that $G^{(0)}(d)$ decays exponentially
beyond the curvature $1/\kappa$, even for $\xi^{-1}=0$ . However,
what matters is the eigenbasis of the correlation function, where
$G$ becomes scale invariant at $T_{c}$. The decay of the correlation
function in real-space will, however, have impact on the $1/N$ corrections
of the critical exponents. Defining a critical exponent $\nu$ through
${\xi}$, we find from (\ref{eq:xiNearTc}) and (\ref{eq:xidef})
the exponent 
\begin{eqnarray}
\xi\sim(T-T_{c})^{-1}\rightarrow\nu=1.
\end{eqnarray}
According to (\ref{chiMomentum}) the zero-momentum susceptibility
is 
\begin{eqnarray}
\chi_{q=0}=\frac{1}{\mu+\kappa^{2}}
\end{eqnarray}
and by using (\ref{eq:xiNearTc}) we find $\gamma=2$. \\
We emphasize here explicitly the fact that both exponents are not
mean-field exponents. The latter are given by $\nu_{\text{MF}}=\frac{1}{2}$
and $\gamma_{\text{MF}}=1$.

\section{Corrections to critical exponents}

\label{sec:ExponentsCorrections}

\begin{figure}[b]
\includegraphics[width=1\linewidth]{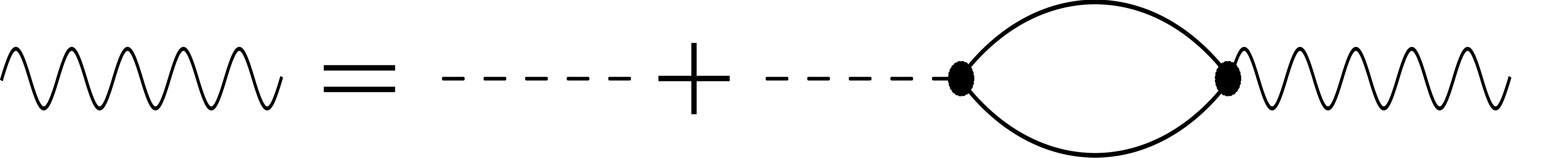}
\protect\caption{\label{fig:2} Dyson equation for the screened interaction $D(p)$}
\end{figure}

Corrections to the critical exponents $\eta$ and $\gamma$ are found
by inspecting the self-energy. In flat space this calculation is described
in \cite{PhysRevA.7.2172} and we find that the general procedure
carries over to hyperbolic space. This procedure consists in first
determining how a correction to a critical exponent would manifest
itself in the self-energy and calculating the order $O(1/N)$ diagrams
to see if such contributions are present.\\
 We start with the action (\ref{eq:action}). The Dyson equation in
real-space reads \begin{widetext} 
\begin{eqnarray}
G(\bm{r}_{P},\bm{r}_{R})=G^{(0)}(\bm{r}_{P},\bm{r}_{R})+\int d\bm{r}_{Q}\int d\bm{r}_{Q'}\ G^{(0)}(\bm{r}_{P},\bm{r}_{Q})\Sigma(\bm{r}_{Q},\bm{r}_{Q'})G(\bm{r}_{Q'},\bm{r}_{R})
\end{eqnarray}
\end{widetext} and is converted to 
\begin{eqnarray}
G^{-1}(p)=G^{(0)-1}(p)+\Sigma(p)
\end{eqnarray}
by application of the convolution theorem. We follow \cite{PhysRevA.7.2172}
in rewriting this equation as 
\begin{eqnarray}
G^{-1}(p)=\left(G^{(0)-1}(p)+\Sigma(0)\right)+\Sigma(p)-\Sigma(0)
\end{eqnarray}
and redefining the new inverse bare Green's function $G^{0-1}(p)$
to be the first term, i.e. 
\begin{eqnarray}
G^{0-1}(p)=\mu_{0}+\Sigma(0)+\kappa^{2}+p^{2}=\mu+\kappa^{2}+p^{2}.
\end{eqnarray}
This has the advantage that at $T=T_{c}$ the `mass' $\mu+\kappa^{2}$
of the bare propagator vanishes. As a consequence of this redefinition,
self-energy insertions in diagrams now take the form $\Sigma(p)-\Sigma(0)$
instead of $\Sigma(p)$. \\
 The large-$N$ structure of the model allows us to restrict ourselves
to a small number of diagrams. The calculated corrections will be
exact to order $1/N$. The coupling constant $u$ in the action is
multiplied by a factor $1/N$. Due to the presence of $N$ fields,
there is a summation over the field index at every $(\phi\cdot\phi)^{2}$
interaction vertex. We represent this interaction term by a dashed
line. On the other hand, a summation over the field index at every
vertex produces a factor $N$. Thus the series of bubbles connected
by $-u/N$ interaction lines, as shown in Figure \ref{fig:2}, are all of order $1/N$ and need to be
included for consistency. We denote this sum by a wiggly line and
use the symbol $D(p)$. It satisfies the relation 
\begin{eqnarray}
D(p)=-\frac{u/N}{1+\frac{u}{2}\Pi(p)}\approx-\frac{2}{N\Pi(p)}\label{eq:DfromPi}
\end{eqnarray}
where $\Pi(p)$ is the polarization operator. The approximation comes
from the large-$u$ limit, which we will be considering from here
on. In real-space $\Pi$ is given by 
\begin{eqnarray}
\Pi(\bm{r},\bm{r}')=G^{0}(\bm{r},\bm{r}')^{2}=\frac{\kappa^{2}}{4\pi^{2}}\frac{e^{-2\sqrt{\mu+\kappa^{2}}d}}{\sinh^{2}\kappa d},
\end{eqnarray}
where $d$ is the geodesic distance between $\bm{r}$ and $\bm{r}'$.
Using (\ref{eq:fourierComp}) we find 
\begin{eqnarray}
\Pi_{p}(\mu)=-\frac{1}{2\pi p}\mathrm{Im}\left[\psi\left(\frac{1}{2}-\frac{ip}{2\kappa}+\frac{\sqrt{\mu+\kappa^{2}}}{\kappa}\right)\right],\label{eq:polOp}
\end{eqnarray}
where $\psi(z)\equiv\frac{d}{dz}\log{\Gamma(z)}$ is the digamma function.
With (\ref{eq:DfromPi}) we find 
\begin{eqnarray}
D_{p}(\mu)=\frac{2\pi p}{N}\frac{1}{\mathrm{Im}\left[\psi\left(\frac{1}{2}-\frac{ip}{2\kappa}+\frac{\sqrt{\mu+\kappa^{2}}}{\kappa}\right)\right]}.\label{eq:screenedPot}
\end{eqnarray}

\begin{figure}[b]
\centering{}\includegraphics[width=0.6\linewidth]{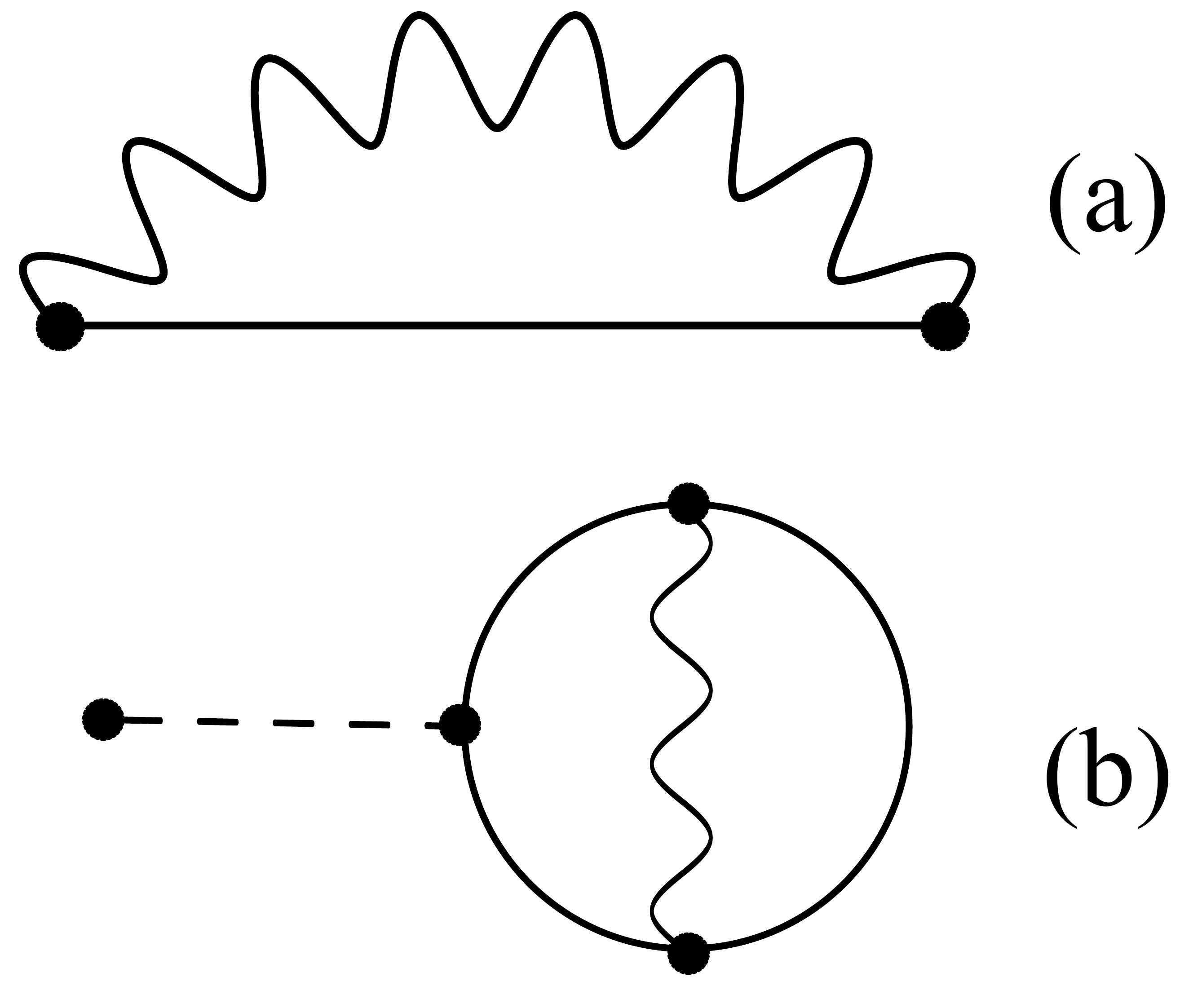}
\protect\caption{\label{fig:3} Relevant $O(1/N)$ diagrams. In flat space, diagram
(a) contributes to $\eta$ and diagram (b) to $\gamma$.}
\end{figure}

We present the calculation of the correction to $\eta$ in detail.
The calculation of the correction to $\gamma$ is much more tedious
and is only sketched. The result in both cases is the absence of any
corrections due to the regularizing character of finite curvature.

\subsection{Order $\mathcal{O}(1/N)$ correction of $\eta$}

\begin{figure}[b]
\centering{}\includegraphics[width=1\linewidth]{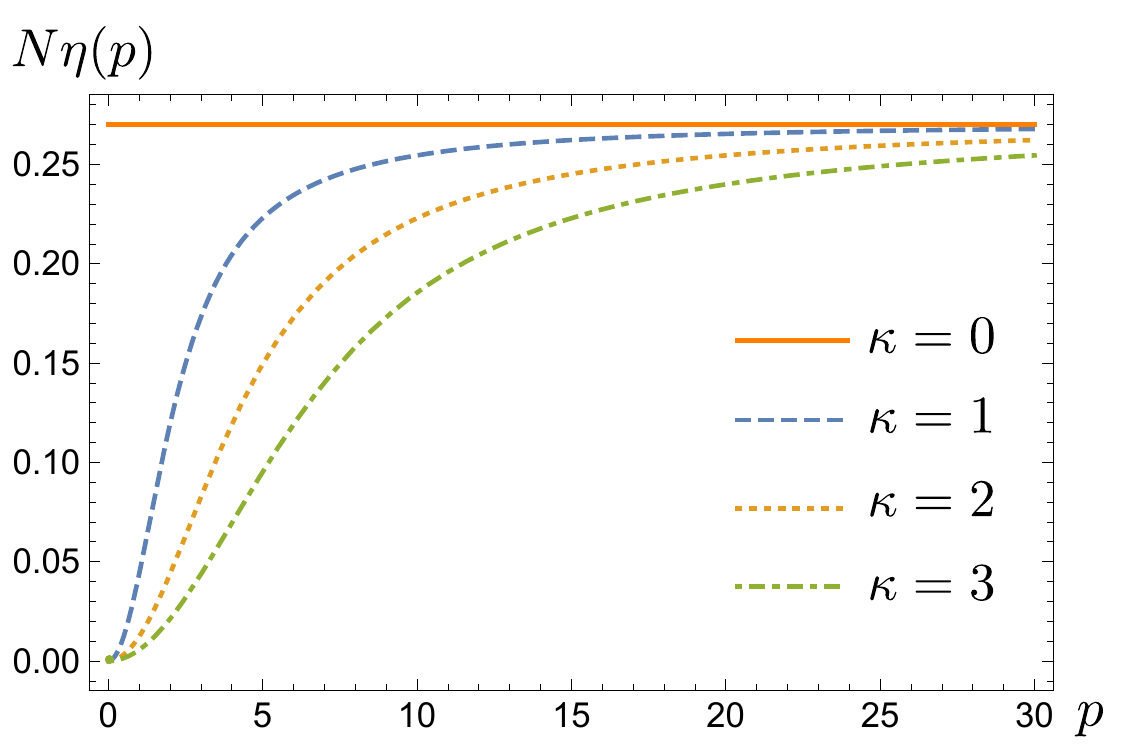}\protect\caption{\label{fig:4} Here we have defined a function $\eta(p)\equiv\frac{d}{d\log p}\left[\frac{\Sigma(p)-\Sigma(0)}{p^{2}}\right]$.
In flat space $\eta(p)=\frac{8}{3\pi^{2}N}$, whereas in hyperbolic
space it is always regularized by curvature and tends to $0$ at long
length-scales.}
\end{figure}

As we have seen $\eta=0$ at lowest order in $1/N$. We now determine
the $1/N$ correction to this result. We have defined $\eta$ in (\ref{eq:defEta}).
Such a correction would manifest itself in the self-energy. For large-$N$
this critical exponent can be expanded and reads 
\begin{eqnarray}
G(p)=\frac{\kappa}{4\pi}\frac{\Lambda^{-\eta}}{p^{2-\eta}}=\frac{\kappa}{4\pi}\frac{\Lambda^{-\eta}}{p^{2}-\eta\ p^{2}\log p},
\end{eqnarray}
i.e. a correction would lead to a $p^{2}\log p$ term in the self-energy
and could be found as the coefficient of such a term. In flat space
there is indeed such a correction. We will now show that this $p^{2}\log p$
term of flat space is regularized in hyperbolic space. The polarization
operator (\ref{eq:polOp}) at the critical point $\mu=-\kappa^{2}$
becomes 
\begin{eqnarray}
\Pi(p)=\frac{1}{4p}\tanh\left(\frac{\pi}{2\kappa}p\right).
\end{eqnarray}
In flat-space the $p^{2}\log p$ contribution is produced by the diagram
in Figure \ref{fig:3}a. To write down this term we take the real-space
form of $D(p)$, which is obtained from (\ref{eq:DfromPi}) with (\ref{eq:fourierComp})
\begin{eqnarray}
D(d_{PQ}) & = & \frac{8\kappa^{4}\cosh(\kappa d_{PQ})}{N\pi^{2}\sinh^{4}(\kappa d_{PQ})}.
\end{eqnarray}
The self-energy in real-space is obtained by multiplying $D$ with
the bare Green's function 
\begin{eqnarray}
\Sigma(d_{PQ}) & = & D(d_{PQ})G^{(0)}(d_{PQ})\\
 & = & \frac{8\kappa^{5}}{N\pi^{3}}\frac{\cosh(\kappa d_{PQ})}{\sinh^{5}(\kappa d_{PQ})}.\label{self-energy}
\end{eqnarray}
Using again formula (\ref{eq:fourierComp}), we can write this in
momentum space as\\
 
\begin{eqnarray}
\Sigma(p) & = & \frac{8(\kappa^{2}+p^{2})}{3N\pi^{2}}\times\label{eq:selfEnergyeta}\\
 &  & \left[\text{Re}\psi\left(-\frac{3}{2}+i\frac{p}{2\kappa}\right)+\frac{2/3\ p^{2}}{9\kappa^{2}+p^{2}}+\gamma-\frac{5}{2}\right],\nonumber 
\end{eqnarray}
where $\gamma$ is the Euler-Mascheroni constant. Taking the flat-space
limit $\kappa\rightarrow0$ with fixed $p$, we obtain the asymptotic
relation 
\begin{eqnarray}
\Sigma(p)\sim\frac{8}{3\pi^{2}N}p^{2}\left(\log\frac{p}{2\kappa}+\gamma-\frac{11}{6}\right).
\end{eqnarray}
The appearance of $\kappa$ in this formula is owed to the fact that
we measure all momenta in units of $\kappa$. \\
In the opposite regime, where $p$ tends to $0$ for fixed curvature,
we have instead of a logarithmic divergence the finite value 
\begin{eqnarray}
\Sigma(p)\sim\frac{8\kappa^{2}}{3\pi^{2}N}\left(\frac{17}{6}-4\log2-\gamma\right).
\end{eqnarray}
This regularizing behavior of the finite curvature is shown in Figure
\ref{fig:4}, where we defined a quantity $\eta(p)\equiv\frac{d}{d\log p}\left[\frac{\Sigma(p)-\Sigma(0)}{p^{2}}\right]$,
which in flat space would yield a finite $\eta$. In hyperbolic space
at sufficiently small $p$, i.e. long length-scales, the log behavior
of the self-energy (\ref{eq:selfEnergyeta}) is cut off and $\eta(p)$
is suppressed to $0$.

\subsection{Order $\mathcal{O}(1/N)$ correction of $\gamma$ }

The exponent $\gamma$ is found from the divergence of the susceptibility
at $p=0$. We find at zero-momentum for the full Green's function
\begin{eqnarray}
G^{-1}\left({\mu(T)}\right) & = & \mu_{0}(T)+\kappa^{2}+\Sigma\left(\mu(T)\right)\nonumber \\
 & = & \mu(T)+\kappa^{2}.
\end{eqnarray}
In the following all calculations will be for zero external momenta,
thus we have suppressed the momentum arguments. Subtracting from this
equation the same equation evaluated at $T=T_{c}$, we find 
\begin{eqnarray}
\mu(T)+\kappa^{2}-\Sigma\left(\mu(T)\right)+\Sigma\left(0\right)=\mu_{0}(T-T_{c}).
\end{eqnarray}
We have found $\gamma=2$ at lowest order in section \ref{subsec:nuGamma}.
Writing $1/\gamma=1/2-\Delta$ and using the definition of the exponent
via $\mu_{0}(T-T_{c})\sim\left(\mu(T)+\kappa^{2}\right)^{1/\gamma}$
for $T$ near $T_{c}$, we obtain 
\begin{eqnarray}
-\Sigma\left(\mu(T)\right)+\Sigma\left(0\right) & \sim & \sqrt{\mu(T)+\kappa^{2}}\\
 &  & -\Delta\sqrt{\mu(T)+\kappa^{2}}\log\left({\mu(T)+\kappa^{2}}\right)\nonumber 
\end{eqnarray}
valid near $T\gtrsim T_{c}$. The first term on the right-hand side
is an $O(1)$ term and was already obtained in section {\ref{sec:exponents}}.
It is produced by a diagram, which is obtained from diagram \ref{fig:3}b
by removing the internal wiggly line. \\
 There are two $O(1/N)$ diagrams that have to be considered in computing
the correction $\Delta$, shown in Figures \ref{fig:3}a and \ref{fig:3}b.
In flat space it can be shown that the diagram in Figure \ref{fig:3}a
only gives a $\sim\mu\log\mu$ correction, whereas the diagram in
Figure \ref{fig:3}b in fact yields a finite $\Delta$. We shall see
now that in hyperbolic space neither diagram yields a contribution
to $\Delta$, as both logarithmic divergences are regularized by $\kappa$.

\subsubsection{Diagram (a)}

We begin with the diagram in Figure \ref{fig:3}a. We denote this
self-energy part by $\Sigma_{a}(\mu)$. Using eq. (\ref{eq:zeroMomConv})
we find \begin{widetext} 
\begin{eqnarray}
\Sigma_{a}(\mu)-\Sigma_{a}(0)=\frac{1}{8\pi\kappa}\int\limits _{-\Lambda}^{\Lambda}dq\int\limits _{-\Lambda}^{\Lambda}dp\left[G_{p}(\mu)D_{q}(\mu)-G_{p}(0)D_{q}(0)\right]\frac{pq}{\cosh^{2}\left(\frac{\pi}{2\kappa}(p-q)\right)}.
\end{eqnarray}
\end{widetext} The hyperbolic cosine effectively cuts off contributions
with $|p-q|\gg\kappa$. Hence the $q$-integral may be approximated
by limiting the integration to this region. Insertion of $G_{p}(\mu)$
and $D_{q}(\mu)$ from (\ref{eq:bareGreen}) and (\ref{eq:screenedPot})
and subsequent expansion in $\mu$ results in a large number of elementary
integrals. All logarithmic terms stemming from these integrals are
of the form $\log(c\kappa^{2}+\mu)$, where $c$ is either $1$ or
$4$. In other words, for finite $\kappa$ no terms proportional to
$\log\mu$ are present.

\subsubsection{Diagram (b)}

Similarly the $\sqrt{\mu}\log\mu$ divergence of the diagram \ref{fig:3}b
in flat-space is regularized. The self-energy expression corresponding
to this diagram is obtained by noticing that diagram \ref{fig:3}b
is the result of attaching to the self-energy in \ref{fig:3}a two
legs of the interaction vertex. It is correspondingly given by \begin{widetext}
\begin{eqnarray}
\Sigma_{b}(\mu) & = & \frac{1}{8\pi\kappa}\int dq\int dq'\ qq'\ G_{q}(\mu)^{2}\left[\Sigma_{a}(q',\mu)-\Sigma_{a}(0,\mu)\right]\frac{1}{\cosh^{2}{\frac{\pi(q-q')}{2\kappa}}}\nonumber \\
 & \approx & \frac{\kappa^{2}}{2\pi^{2}}\int dq'\ \frac{q'^{2}}{[(q'-2\kappa/\pi)^{2}+\mu+\kappa^{2}][(q'+2\kappa/\pi)^{2}+\mu+\kappa^{2}]}\left[\Sigma_{a}(q',\mu)-\Sigma_{a}(0,\mu)\right]\label{eq:sigmaGammaB}
\end{eqnarray}
\end{widetext} where the same approximation as before has been made.
In the integrand the momentum-dependent self-energy $\Sigma_{a}(q,\mu)$
is required. According to (\ref{eq:finMomConv}) this is given by
\begin{widetext} 
\begin{equation}
\Sigma_{a}(q',\mu)=\frac{1}{4\pi^{2}q'\kappa}\int\limits _{-\Lambda}^{\Lambda}dp'\int\limits _{-\Lambda}^{\Lambda}dp\left[G_{p}(\mu)D_{p}'(\mu)-G_{p}(0)D_{p}'(0)\right]pp'\tanh\left(\frac{p+p'-q'}{2\kappa}\pi\right).\label{eq:sigmaGammaA}
\end{equation}
\end{widetext} Inside the $p$-integral we approximate the $\tanh$-function
in the region $|p+p'-q|<\frac{2\kappa}{\pi}$ by its argument and
outside this region by the sign-function. Then the $p$-integral may
be carried out without a cutoff and we are left with a {$p'$-integral}.
Insertion of (\ref{eq:sigmaGammaA}) into (\ref{eq:sigmaGammaB})
and integration over $q'$ results in \begin{widetext} 
\begin{eqnarray}
\Sigma_{b}(\mu)=\frac{2\pi^{2}}{\kappa\lambda}\int\limits _{0}^{\Lambda}dq'\frac{q'^{2}}{\mathrm{Im}\psi\left(\frac{1}{2}-\frac{iq'}{2\kappa}+\frac{\lambda}{\kappa}\right)}\log\frac{(q'-2\kappa)^{2}+4\lambda^{2}}{(q'+2\kappa)^{2}+4\lambda^{2}}+R(\kappa).\label{eq:sigma_b}
\end{eqnarray}
\end{widetext} where $R(\kappa)$ denotes terms that tend to $0$
with $\kappa\rightarrow0$. The first term on the right-hand side
reproduces for $\kappa=0$ fully the flat space formula for $\Sigma_{b}$.
This self-energy contains the $\sqrt{\mu}\log\mu$ term. For finite
$\kappa$, however, the integral in (\ref{eq:sigma_b}) is fully regularized
and a $\sqrt{\mu}\log\mu$ term is avoided. Thus, we conclude that
no singular correction to the $\mu$ dependence of the self-energy
emerges, i.e. the exponent $\gamma$ is also unchanged compared to
the leading order $1/N$ expression given above.

\section{Discussion}

\label{sec:discussion}

The aim of this paper was to investigate critical phenomena in hyperbolic
space. Our key finding is that for a $\bm{\phi}^{4}$-model embedded
in hyperbolic space a new fixed point emerges at finite curvature
$\kappa$. If $\kappa>0$ the critical exponents are governed by the
strong curvature limit. Interestingly, these exponents are given by
leading order terms of the $1/N$ expansion. Thus, while the numerical
values of the exponents are now simpler, they continue to obey hyperscaling
below the upper critical dimension. The physical state in the symmetry-broken
regime is characterized by an unusual magnetization texture. This
texture consists of regions of size of the order of the radius of
curvature $1/\kappa$ where the vector $\bm{\phi}$ has nearly uniform
direction. Beyond this region the finite value of the curvature starts
to play an important role, since a global direction in hyperbolic
space is not a well-defined concept. It is therefore not possible
to establish a uniform direction of the magnetization vector. In fact,
as we have illustrated in Figure \ref{HypTriangle}, the parallel
transport of a local direction from region $A$ to $B$ and then from
$B$ to $C$ is not the same as the direct parallel transport from
$A$ to $C$. It is this lack of transitivity which is the origin
of the resulting magnetization texture. \\
The fact that the $\kappa\neq0$ values of exponents are different
from the flat space $\kappa=0$ limit may be understood using standard
crossover arguments as we now show using the example of magnetic susceptibility.
Let $f(t,h,\kappa)$ be the singular part of the free energy density,
where $t\propto(T-T_{c})/T_{c}$ measures the distance to the critical
point and $h$ is the external field. Then the following scaling transformation
holds 
\begin{equation}
f(t,h,\kappa)=b^{-3}f\left(b^{1/\nu_{f}}t,b^{y_{f}}h,b\kappa\right),
\end{equation}
with exponents $\nu_{f}$ for the correlation length and scaling dimension
of the conjugate field $y_{f}=\beta_{f}\delta_{f}/\nu_{f}$ that refer
to the flat space ($\kappa=0$) limit. The curvature is a relevant
perturbation with positive scaling dimension, i.e. the infrared behavior
is governed by the infinite curvature fixed point, where all scales
(except of course for the inverse ultraviolet cut-off) are larger
compared to the radius of curvature. Performing the second derivative
with respect to the conjugate field, we obtain the scaling expression
for the order parameter susceptibility: 
\begin{eqnarray}
\chi(t,\kappa) & = & b^{\gamma_{f}/\nu_{f}}\chi(b^{1/\nu_{f}}t,b\kappa)\nonumber \\
 & = & t^{-\gamma_{f}}\Phi\left(\frac{\kappa}{t^{\nu_{f}}}\right).
\end{eqnarray}
In the flat space limit $\kappa=0$, the scaling function behaves
as $\Phi(x\rightarrow0)\rightarrow\text{const.}$ and we recover the
flat space results. On the other hand, our above analysis implies
that for large argument $\Phi(x\gg1)\propto x^{-\phi}$ holds with
crossover exponent 
\begin{equation}
\phi=\frac{\gamma-\gamma_{f}}{\nu_{f}}=\frac{24}{N\pi^{2}}.
\end{equation}
Here $\gamma$ is the susceptibility exponent of the hyperbolic space
obtained above. Thus, we find $\chi(t,\kappa)\propto\kappa^{-\phi}t^{-\gamma}$.
The behavior $\kappa^{-\phi}$ is, at the considered order, fully
consistent with the $\phi\log(\kappa)$ behavior that occured in our
explicit analysis. We have calculated the critical exponents $\eta,\gamma$
and $\nu$ at lowest order in $1/N$ and found that these are identical
to the exponents in flat three-dimensional space at lowest order.
For $\eta$ and $\gamma$ we showed that $\textrm{O}(1/N)$ corrections
are absent. As our calculations show, the reason for this absence
is the fact that correlations are exponentially decaying beyond the
radius of curvature even at the critical point. The lowest order values
of the exponents are computed from local quantities, which are oblivious
to the finite curvature, whereas the higher-order corrections are
determined through integration over the whole of hyperbolic space,
wherein the finite curvature serves to cut off the long-wavelength
fluctuations. For this reason, we may also surmise the absence of
corrections to the other critical exponents. It is moreover plausible
to assume for the same reason that higher-order corrections to the
exponents will also be absent in the $1/N$-expansion. Thus we conjecture
that the critical exponents we found are correct to all orders in
$1/N$.\\
An interesting question is how our results are modified for dimensions
$d$ different from $3$. The Laplacian in $d\neq3$ dimensions is
still gapped. The only modification in our lowest order $1/N$ calculations
of the critical exponents would be a change of the integration measure
in (\ref{eq:nu1}), from $p^{2}$ to $p^{d-1}$, multiplied by a numerical
factor. However, this leads again the same saddle point equations
as in flat space. Thus we can make the stronger statement that all
critical exponents in hyperbolic space are just the leading order
$1/N$ exponents of flat space. In particular, we have $\nu=\frac{1}{d-2}$
and $\gamma=\frac{2}{d-2}$ for $d\leq4$ and mean-field exponents
for $d>4$. The upper critical dimension is $d=4$ even for finite
$\kappa$. \\
In summary,  we conclude that the description of many-particle systems
in hyperbolic space is a promising avenue to investigate uniform frustration
and non-trivial critical behavior within one theoretical approach.
\begin{acknowledgments}
We are grateful for discussions with \mbox{Ulf} \mbox{Briskot},
\mbox{Vladimir} \mbox{Dobrosavljevi}\'{c}, \mbox{Rafael} \mbox{Fernandes},
\mbox{Eduardo} \mbox{Miranda}, \mbox{Peter}\inputencoding{latin1}{
}\inputencoding{latin9}\mbox{Wolynes} and \mbox{Jan} \mbox{Zaanen}. 
\end{acknowledgments}

\bibliographystyle{phreport}

\end{document}